\documentclass[12pt]{iopart}
\usepackage{iopams} 
\usepackage[usenames,dvipsnames]{xcolor} 
\begin{document}

\title[Can test fields destroy the event horizon in the Kerr-Taub-NUT spacetime?]{Can test fields destroy the event horizon in the Kerr-Taub-NUT spacetime?}

\author{Koray D\"{u}zta\c{s}}

\address{Physics Department, Eastern Mediterranean  University, Famagusta, North Cyprus, via Mersin 10, Turkey}
\ead{koray.duztas@emu.edu.tr}
\vspace{10pt}

\begin{abstract}
In this work we investigate if the interaction of the Kerr-Taub-NUT spacetime with test scalar and neutrino fields can lead to the destruction of the event horizon. It turns out that both extremal and nearly extremal black holes can  be destroyed by scalar and neutrino fields if the initial angular momentum of the spacetime is sufficiently large relative to its mass and NUT charge. This is the first example in which a classical field  satisfying the null energy condition can actually destroy an extremal black hole. For scalar fields, the modes that can lead to the destruction of the horizon are restricted to a narrow range due to superradiance. Since superradiance does not occur for neutrino fields, the destruction of the horizon by neutrino fields is generic, and it cannot be fixed by backreaction effects. We also show that the extremal black holes that can be destroyed by scalar fields correspond to naked singularities in the Kerr limit, in accord with the previous results which imply that extremal Kerr black holes cannot be destroyed by scalar test fields.
\end{abstract}

%
\vspace{2pc}
\submitto{\CQG}
%
%
%

\section{Introduction}
The Kerr-Taub-NUT (Newman-Unti-Tamburino) (KTN) space-time is an analytic type D vacuum solution of the Einstein equations  which was first derived by Demia\'{n}ski and Newman~\cite{dem}. The metric can be expressed in Kerr-like coordinates: 
\begin{eqnarray}
d s^2
   &=&\frac{1}{\Sigma}(\Delta -a^2\sin^2\theta)d t^2
    -\frac{2}{\Sigma}[\Delta A -a(\Sigma +a A)\sin^2\theta]d t
d \phi \nonumber \\ 
    &&-\frac{1}{\Sigma}[(\Sigma +a A)^2\sin^2\theta-A^2\Delta] d \phi^2
    -\frac{\Sigma}{\Delta}d r^2 -\Sigma d\theta^2
\label{metricktn}
\end{eqnarray}
where $m$ is the gravitational mass, $-\ell$ is the gravitomagnetic monopole moment or the NUT charge, and $a$ is the Kerr parameter. $\Sigma$, $\Delta$ and $A$ are defined by
\begin{eqnarray}
&\Sigma& = r^2 +(\ell +a \cos \theta)^2,\ 
\Delta = r^2-2Mr-\ell^2 + a^2 \nonumber \\ 
&A& = a \sin^2\theta -2\ell\cos\theta 
\end{eqnarray}
Though the metric (\ref{metricktn}) resembles that of Kerr, the singularities at $\theta=0,\pi$ are not the usual degeneracies of the two sphere.
The same problem was previously encountered in Taub-NUT space-time. For this case, Misner showed that these singularities become the degeneracies of the spherical coordinates on the three sphere by imposing a certain identification on the metric~\cite{misner}. The same identification was used by Miller in his global analysis of the KTN space-time~\cite{miller}. 
\begin{equation}
(\phi,t)=\left( \phi + (n+m)2\pi, t + (n-m)4 \ell \pi \right)
\label{iden}
\end{equation}  
where  $n,m$ are integers. Miller considers the case where the roots of $\Delta(r=r_{\pm})=0$ are real and distinct  so that there exists an inner and an outer event horizon with coordinates $r_{\pm}$. Three disjoint regions are defined where the metric is non-singular except the possible curvature singularity: $-\infty < r <r_-$, $r_- < r < r_+$, and $r_+<r<\infty$.  Each region is a connected Lorentz manifold with a metric obtained by imposing  the identification (\ref{iden}) on (\ref{metricktn}). In the global analysis of the KTN space-time, Miller constructs extensions of $M_1$, $M_2$, and $M_3$ that are analytic on $S^3 \times R$. For $a>\vert \ell \vert$, $\Sigma$ can be zero, and there exists a curvature singularity  inside the event horizon of the KTN black hole at $r=0$ and $\theta= \arccos [(- \ell )/a]$.  For $a<\vert \ell \vert$, $\Sigma$ is positive definite, and there is no curvature singularity in the spacetime similar to the Taub-NUT case. In the region $r_+<r<\infty$ the radial coordinate $r$ is spacelike, and hypersurfaces of constant $r$ are timelike 3-spheres with closed timelike curves.

The identification (\ref{iden}) apparently forces periodicity in the time coordinate, which renders the Taub-NUT and KTN space-times unphysical as they admit closed time-like curves. To avoid closed time-like curves, Bonnor suggested to impose only part  of Misner's identification \cite{bonnor}, and interpreted the singularity at $\theta=\pi$ as a massless source of angular momentum, which cannot be covered in an extension.  A maximal analytic extension of the KTN metric with Bonnor's identification was also obtained in the global analysis by Miller. Despite its unpleasant causal behaviour, particle motion on the background of the KTN space-time is extensively studied to evaluate the possible effects of gravitomagnetic monopoles if they exist~\cite{ktn1,ktn2,ktn3,ktn4,ktn5,ktn6,ktn7}. 

In general relativity, one of the ways to identify a spacetime is to study its perturbations. For example, if there exists exponentially growing modes for massless fields, the spacetime is unstable. It will not be able to return to its original state after perturbation. In this field the most outstanding contribution was given by Teukolsky, who decoupled and separated the wave equations for all massless test fields on Kerr background, and combined them into a single master equation parametrised by the spin parameter~\cite{teuk1}. Teukolsky used Newman-Penrose two-spinor formalism to decouple the wave equations in a general context valid for all type D vacuum space-times, then he proceeded to apply separation of variables on Kerr background.  A new version of Teukolsky's master equation was derived in \cite{teukbini}. Similar analysis were performed on different type D vacuum backgrounds including  the Taub-NUT~\cite{bini0},  the Kerr-Taub-NUT~\cite{bini}, and the C metric~\cite{kofron}.

The cosmic censorship conjecture (CCC) asserts that all the singularities that form in gravitational collapse are hidden behind the event horizons of black holes~\cite{ccc}, so that no external observer can be in causal contact with a singularity.  Causality violations occur in spacetimes with closed timelike curves. Closed timelike curves are also present in the Kerr spacetime, which is the best candidate to describe an astrophysical black hole. In Kerr case, the closed timelike curves are confined inside the event horizon which disconnects them from the rest of the universe. Hence, the event horizon hides the naked singularity and maintains the causal behaviour of the spacetime (see e.g.~\cite{joshi1}).  However, there is no direct connection between causality violation and occurrence of naked singularities. There exists dynamical spacetimes such as the L\'{e}maitre-Tolman-Bondi dust collapse models which admit naked singularities but no closed timelike curves~\cite{joshi2}.

As a concrete proof of CCC has been elusive, the stability of event horizons in the interactions of black holes with test particles and fields has become one of the most intriguing problems in black hole physics. If the event horizon of a black hole can be destroyed, we end up with a naked singularity visible to outside observers. Therefore  this problem is closely related to CCC, though it is not identical. In the case of test particles one usually analyses the geodesic motion, and in the case of fields the analysis consists of a scattering problem.  In this respect many thought experiments involving the scattering of waves were constructed in the Kerr and Kerr-Newman space-times~\cite{semiz,toth1,emccc,overspin,superrad,sh,duztas,toth,natario,duztas2,mode}. It turns out that there exists a generic violation of CCC due to neutrino fields~\cite{duztas,toth,natario}. (This is often confused with the attempts to destroy the event horizon with a single neutrino. See \cite{mode} for a general discussion) In similar attempts to overspin or overcharge asymptotically AdS black holes, the authors concluded that CCC cannot be violated~\cite{btz1,btz2,vitor,zhang,rocha,gwak1,gwak2}. However,  we have  recently shown that it is possible to overspin a Ba\~nados-Teitelboim-Zanelli (BTZ) black hole into a naked singularity using both test particles and fields~\cite{btz}.

In this work we consider a KTN black hole with a curvature singularity at the center $(a>\vert \ell \vert)$. We test the stability of the event horizon as massless scalar and neutrino fields studied by Bini {\it et al}. scatter off the  black hole. The main difference between scalar and neutrino fields is the occurrence of superradiance which plays a key role in scattering problems. We apply the thought experiment both to extremal and nearly extremal black holes. Finally we comment on the results regarding their relevance with previous works.

\section{Can Test Fields Destroy The Event Horizon?}
In thought experiments to test the stability of event horizons against test particles, one first finds the upper limit for the angular momentum or charge of the test particle so that it crosses the horizon. Naturally, there exists a lower limit so that the test particle can overspin or overcharge the black hole into a naked singularity.  In the first of these experiments, Wald found that particles with enough charge or angular momentum, to  overcharge/overspin  an extremal Kerr-Newman black hole into a naked singularity, do not cross the horizon to be captured by the black hole~\cite{wald74}. Later  Hubeny developed an alternative approach, and showed that if we start with a nearly extremal black hole instead of an extremal one, it is possible to overcharge a Reissner-Nordst\"{o}m black hole~\cite{hubeny}. Similar analysis were performed for Kerr-Newman and Kerr-Sen black holes~\cite{saa,gao,kerrsen}. Adopting the same approach, Jacobson and Sotiriou found a range of possible values for the angular momentum of a test particle, so that it crosses the horizon and overspins a Kerr black hole~\cite{jacobsot}. We applied a similar procedure for bosonic test fields interacting with Kerr black holes~\cite{overspin}. For bosonic test fields there exists a minimum value for the frequency $\omega$ so that the particle is absorbed by the black hole. If the frequency of the incident field is  below this critical value, superradiance occurs. Superradiance can roughly be defined as the amplification of waves as they scatter off black holes. The minimum value of the frequency corresponds to the maximum value of the fields contribution to the angular momentum of the background spacetime ($\delta J$), analogous to the particle case. On the other hand, the minimum value for $\delta J$ (or the maximum value for $\omega$) is determined by requiring that the horizon is destroyed at the end of the interaction. We found that there exists a range of frequencies that can destroy the horizon if we start with a nearly extremal black hole. However, it is not possible to overspin an extremal Kerr black hole by using bosonic test fields. Later, this result was generalised by Natario {\it et al}., who proved that test fields satisfying the null energy condition cannot overspin/overcharge extremal Kerr-Newman and Kerr-Newman-anti de Sitter black holes into naked singularities~\cite{natario}. In a recent work involving BTZ black holes we derived analogous results to the Kerr case~\cite{btz}. 

In the thought experiments involving fields, we envisage a test field with frequency $\omega$ and azimuthal wave number $m$ scattering from the black hole. This is possible if the background space-time is stationary and axi-symmetric with corresponding Killing vectors $\partial /\partial t$ and $\partial /\partial \phi$, so that the test fields admit separable solutions in the form 
\begin{equation}
\Psi(t,r,\theta,\phi)=\rme^{-\rmi \omega t}\rme^{\rmi  m\phi}F(r,\theta)
\label{sepa}
\end{equation}
The test field is incident on the black hole from spatial infinity at early times. As $t\to \infty$ the field decays away, and the spacetime is described by the new perturbed parameters.  The contribution of the test fields to mass and angular momentum parameters is related by~\cite{beken}
\begin{equation}
\delta J=(m/\omega)\delta E
\label{beken}
\end{equation}
where $J=Ma$ and $\delta E=\delta M$. In the Kerr case, we started with a nearly extremal spacetime which satisfies $J/M^2=a/M=1- 2\epsilon^2$, where $a$ and $M$ are the usual angular momentum and mass parameters. We derived the maximum value for the frequency of the incoming field by requiring that the horizon is destroyed at the end of the interaction. We showed that the maximum value for the frequencies is slightly larger than the limiting frequency for superradiance ($\omega_{\rm{sl}}$). Therefore the frequencies in the range $\omega_{\rm{sl}} < \omega < \omega_{\rm{max}}$ can be used to overspin nearly extremal black holes. However, this range vanishes if the black hole is initially extremal.

In this work, we apply a similar procedure to test the stability of the event horizon in the KTN spacetime. The test fields in the KTN space-time satisfy the master equation derived by Bini {\it et al}. in \cite{bini}. They admit separable solutions of the form (\ref{sepa}). One can envisage a test field incident on the KTN black hole, with frequency $\omega$ and azimuthal wave number $m$, which is partially absorbed  and partially reflected back to infinity. After the interaction, the test field decays away, leaving behind another KTN spacetime with new parameters.  Since the incident waves  carry energy and angular momentum, the interaction  leads to perturbations in mass ($M$) and angular momentum parameters ($a$) of the background space-time.  The contribution of the test fields to mass and angular momentum parameters is related by (\ref{beken}). Initially the space-time satisfies $M^2+\ell^2-a^2>0$ or equivalently
\begin{equation}
\delta \equiv M^2+\ell^2-\frac{J^2}{M^2}>0
\label{crit}
\end{equation}
Before the interaction there exists an inner and an outer event horizon at $r_{\pm}=M \pm \sqrt{\delta}$. At the end of the interaction, if the final configuration satisfies (\ref{crit}) the event horizon still exists. However if the final configuration fails to satisfy (\ref{crit}), i.e. $\sqrt{\delta}$ is not a real number, the event horizon has been destroyed. In this thought experiment we search for possible modes of the incoming waves that can drive the space-time beyond extremality. First we are going to drive the maximum value of frequencies by imposing that the horizon is destroyed. Then, for scalar fields, we have to verify that the maximum value of frequencies is greater than the superradiant limit, which is the minimum value to allow the absorption of the incoming field. If $\omega_{\rm{max}} \leq \omega_{\rm{sl}}$, one cannot find a frequency to destroy the event horizon. If  $\omega_{\rm{max}} > \omega_{\rm{sl}}$, the frequencies in the range $  \omega_{\rm{sl}}< \omega < \omega_{\rm{max}}$ can be used to destroy the horizon. Let us start by defining the dimensionless parameters
\begin{equation}
\beta \equiv \frac{\vert \ell \vert}{M}, \quad \alpha \equiv \frac{J}{M^2}=\frac{a}{M}
\label{define}
\end{equation}
Initially the space-time satisfies
\begin{equation}
1+\beta^2 - \alpha^2= \epsilon^2
\label{param}
\end{equation}
With $\epsilon \ll 1$, the space-time is nearly extremal. If the horizon is to be destroyed we demand that the final configuration satisfies
\begin{equation}
(M+\delta E)^2+ \ell^2 < \frac{(J+\delta J)^2}{(M+\delta E)^2}
\label{overspin1}
\end{equation}
For the incoming field we choose that $\delta E \sim M\epsilon$ without violating the test field approximation. $\delta J$ and $\delta E$ are related by (\ref{beken}). Using (\ref{define}), we re-write (\ref{overspin1})
\begin{equation}
M^2\left[ (1+\epsilon)^2 + \beta^2 \right] < \frac{(J + (m/\omega) M\epsilon)^2}{M^2(1+ \epsilon)^2}
\label{overspin2}
\end{equation}
We prefer to express the equations in terms of $\alpha$. For that reason, we use (\ref{param}) to eliminate $\beta$ from (\ref{overspin2}). We  proceed by taking the square root of both sides. After some algebra we find that the event horizon can be destroyed if the frequency of the incoming waves satisfy
\begin{equation}
\omega< \frac{m\epsilon}{M\left[(1+\epsilon)\sqrt{\alpha^2 + 2 \epsilon^2 + 2 \epsilon}- \alpha\right]} \equiv
\omega_{\rm{max}}
\label{critic}
\end{equation}
To destroy the horizon the frequency of the incoming field should satisfy $\omega < \omega_{\rm{max}}$. However, as we pointed out, this condition is not sufficient for the destruction of the horizon if superradiance can occur. For superradiant modes, incoming waves are amplified as they scatter off the black hole. In other words the reflection amplitude is larger than 1, or the absorption probability is negative. The extra energy of the reflected waves is supplied by the rotation of the black hole. Therefore scalar fields should also satisfy $\omega > \omega_{\rm{sl}}$. In ~\cite{bini} Bini {\it et al}. derived that superradiance occurs in the KTN space-time for scalar fields if the frequency of the incoming wave is less than the superradiant limit 
\begin{equation}
\omega_{\rm{sl}}=m\Omega=\frac{ma}{2(Mr_+ + \ell^2)}
\label{superrad}
\end{equation}
where $\Omega=a/2(Mr_+ + \ell^2)$ is the effective angular velocity of the horizon. If we send in scalar waves with $\omega<\omega_{\rm{sl}}$, the wave will be reflected with a larger amplitude, the angular momentum of the black hole will decrease and the main inequality (\ref{crit}) will be reinforced. For that reason, for overspinning to occur we should  also demand that the frequency of the incoming wave is larger than the superradiant limit. 
\begin{equation}
\omega>\omega_{\rm{sl}}=\frac{m\alpha}{2M(\alpha^2 + \epsilon^2 + \epsilon)}
\label{overspin3}
\end{equation}
where we used $r_+=M+\sqrt{M^2 - a^2 + \ell^2}=M(1+\epsilon)$ and $\ell^2=M^2\beta^2=M^2(\alpha^2 + \epsilon^2 -1)$. If the frequency of the incoming scalar field satisfies both (\ref{critic}) and (\ref{overspin3}), i.e. $\omega_{\rm{sl}} < \omega < \omega_{\rm{max}}$, the horizon can be destroyed at the end of the interaction. If  $ \omega > \omega_{\rm{max}}$, the final parameters of the spacetime satisfy (\ref{crit}), and the event horizon is stable. If $ \omega < \omega_{\rm{sl}}$, the field is reflected back with a larger amplitude, (\ref{crit}) remains valid and the event horizon continues to exist. One can find a frequency in the range  $\omega_{\rm{sl}} < \omega < \omega_{\rm{max}}$ to destroy the horizon, provided that $\omega_{\rm{max}}>\omega_{\rm{sl}}$.  In similar works involving Kerr and BTZ black holes we were able to show that $\omega_{\rm{max}}$ is always slightly larger than $\omega_{\rm{sl}}$. However, this is not possible in the KTN spacetime. Demanding that $\omega_{\rm{max}}>\omega_{\rm{sl}}$ will bring an extra condition on the initial parameters of the background spacetime. Using the definitions of $\omega_{\rm{max}}$, and $\omega_{\rm{sl}}$ in  (\ref{critic}) and  (\ref{overspin3}),  the condition  $\omega_{\rm{max}}>\omega_{\rm{sl}}$ is equivalent to 
\begin{equation}
\frac{1}{\epsilon}\left[(1+\epsilon)\sqrt{\alpha^2 + 2 \epsilon^2 + 2\epsilon} - \alpha \right]< \frac{2}{\alpha} \left( \alpha^2 + \epsilon^2 + \epsilon\right) 
\end{equation}
Working up to second order in $\epsilon$, we derive the  condition that should be satisfied by the initial parameters of the spacetime, so that  $\omega_{\rm{max}}$ can be larger than $\omega_{\rm{sl}}$.
\begin{equation}
\alpha^2> \frac{2\epsilon + 2}{3\epsilon + 2} \equiv \alpha^2_{\rm{cr}}
\label{condi2}
\end{equation}
The only way that test fields can destroy the event horizon is to overspin the black hole. For this to occur, the nearly extremal KTN spacetimes parametrized in the form (\ref{param}) should also satisfy the condition (\ref{condi2}); i.e. the initial angular momentum of the spacetime should be sufficiently large relative to is mass and NUT charge. Starting with a nearly extremal spacetime parametrized as (\ref{param}),  we first  require that  the initial parameters of the space-time  satisfies (\ref{condi2}). Otherwise  $\omega_{\rm{max}}$ is smaller than $\omega_{\rm{sl}}$, and one cannot find a frequency for the incoming field to destroy the horizon. The modes that could lead to that could lead to the destruction of the horizon  will be subject to superradiance. If the initial parameters satisfy (\ref{condi2}), then $\omega_{\rm{cr}}>\omega_{\rm{sl}}$, and we send in a scalar field from spatial infinity with frequency in the range $\omega_{\rm{sl}}<\omega<\omega_{\rm{cr}}$, and $\delta E \sim M\epsilon$. At the end of the interaction of the spacetime with this field, the final parameters do not satisfy (\ref{crit}) and the event horizon cannot exist.

For a numerical example, let us chose $\epsilon=0.01$. For the horizon to be destroyed, we first demand that  $\alpha$ is larger than the critical value $\alpha_{\rm{cr}}=0.9975$, which is determined by (\ref{condi2}). If $\alpha<\alpha_{\rm{cr}}$, then  $\omega_{\rm{cr}}<\omega_{\rm{sl}}$, which means that the modes that could destroy the horizon are not absorbed by the black hole. Let us start with a spacetime with $\alpha=1>\alpha_{\rm{cr}}$. The initial parameters of this spacetime are given by $J=M^2 \alpha = M^2$, and $\ell^2= M^2 \beta^2 =M^2(\epsilon^2 + \alpha^2 -1)=M^2 \epsilon^2$. With $\epsilon=0.01$ the initial parameters of the spacetime satisfy
\begin{equation}
\delta_{\rm{in}} \equiv M^2 + \ell^2 -\frac{J^2}{M^2}=0.0001M^2
\label{ex1}
\end{equation}
Since the right hand side of (\ref{ex1}) is positive, before the interaction with the test field, there exists an inner and an outer horizon with spatial coordinates  
\begin{equation}
r_{\pm}=M \pm \sqrt{\delta_{\rm{in}}}=M \pm 0.01M
\end{equation}
For this spacetime  $\omega_{\rm{max}}=0.4963(m/M)$ and $\omega_{\rm{sl}}=0.4950(m/M)$ up to four significant digits. The frequencies in the range $
0.4950(m/M)<\omega<0.4963(m/M)$ can lead to overspinning to destroy the horizon. Let us send in a scalar field from infinity with $\omega = 0.4951 (m/M)$ and $\delta E =M\epsilon=0.01M$. For this field $\delta J= (m/\omega) \delta E = 0.0202 M^2$. In the final configuration, the parameters of the spacetime satisfy
\begin{eqnarray}
\delta_{\rm{fin}} &=& (M+ \delta E)^2 + \ell^2 - \frac{(J+ \delta J)^2}{(M+ \delta E)^2 } \nonumber \\
&=& M^2\left[ (1+0.01)^2 + 0.0001 - \frac{(1+ 0.0202)^2}{(1+0.01)^2 } \right]\nonumber \\ &=& -0.0001 M^2
\label{ex2}
\end{eqnarray}
The right hand side of (\ref{ex2}) is negative. The final parameters of the spacetime does not allow an event horizon to exist. The horizon has been destroyed in the interaction with the scalar field.
\subsection{Neutrino fields to destroy the horizon}
In \cite{bini}, Bini {\it et al}. also derived that superradiance does not occur in the KTN space-time for neutrino fields in analogy with the Kerr case. (Here, neutrino field refers to a massless spin $1/2$ field.) The absorption probability of all modes is positive. If we perturb the KTN space-time with a neutrino field, the range of frequencies that can be used to destroy the horizon is not bounded below by the superradiance limit. All the modes in the range $0< \omega< \omega_{\rm{cr}}$ can be used to destroy the horizon.  That is a generic destruction compared to the case of scalar fields, which is analogous to the generic violation of CCC by neutrino fields in Kerr space-time~\cite{duztas,toth,natario,duztas2,mode}.

Let us consider the spacetime in the previous example with $\alpha=1$ and $\epsilon=0.01$. This time, we send in a neutrino field from infinity. Again $\delta E =M\epsilon=0.01M$ for this field, and we choose  $\omega = 0.25 (m/M)$. Since superradiance does not occur for neutrino fields, we can choose a value below $\omega_{\rm{sl}}$ for the frequency of the incoming field. The contribution of this field to the angular momentum of the spacetime is much larger: $\delta J= (m/\omega) \delta E = 0.04 M^2$. We can calculate $ \delta_{\rm{fin}}$
\begin{eqnarray}
\delta_{\rm{fin}} &=& (M+ \delta E)^2 + \ell^2 - \frac{(J+ \delta J)^2}{(M+ \delta E)^2 } \nonumber \\
&=& M^2\left[ (1+0.01)^2 + 0.0001 - \frac{(1+ 0.04)^2}{(1+0.01)^2 } \right]\nonumber \\ &=& -0.0401 M^2
\label{exneut}
\end{eqnarray} 
The absolute value of  $ \delta_{\rm{fin}}$ is about 400 times larger than that of the previous example with scalar fields. Since there is no superradiance, we can lower the frequency of the incoming field to increase the absolute value $ \delta_{\rm{fin}}$ even further. This is a robust destruction of the horizon compared to the case of scalar fields. 
\subsection{The Extremal case}
In the extremal case, the initial parameters of the space-time satisfy
\begin{equation}
1+\beta^2 - \alpha^2=0
\end{equation}
The condition (\ref{overspin1}) to destroy the horizon  is also valid in this case. We choose $\delta E=M\epsilon^{\prime}$ where $\epsilon^{\prime} \ll 1$ to maintain the test field approximation. The condition that the horizon cannot exist in the final configuration of the  space-time parameters is given by 
\begin{equation}
M^2\left[ (1+\epsilon^{\prime})^2 + \beta^2 \right] < \frac{(J + (m/\omega) M\epsilon^{\prime})^2}{M^2(1+ \epsilon^{\prime})^2}
\label{overspin2ex}
\end{equation}
Equation (\ref{overspin2ex})  has the same structure as equation (\ref{overspin2}) with $\epsilon$ substituted by $\epsilon^{\prime}$. Now we impose that the space-time is extremal by substituting $\beta^2=\alpha^2 - 1$ in (\ref{overspin2ex}), and we derive the critical frequency for the extremal case
\begin{equation}
\omega< \frac{m\epsilon^{\prime}}{M\left[(1+\epsilon^{\prime})\sqrt{\alpha^2 +  \epsilon^{\prime^2} + 2 \epsilon^{\prime}}- \alpha\right]} \equiv
\omega_{\rm{max-ex}}
\label{critic-ex}
\end{equation}
For scalar fields we demand that $\omega_{\rm{max-ex}}>\omega_{\rm{sl-ex}}$ so that the incoming fields are not reflected back with a larger amplitude. The superradiant limit for the extremal case is given by
\begin{equation}
\omega_{\rm{sl-ex}}=\frac{ma}{2(Mr_+ + \ell^2)}=\frac{m}{2M\alpha}
\end{equation}
where we used $r_+=M$, and $\ell^2=M^2\beta^2=\alpha^2-1$ which are valid for the extremal case. The condition $\omega_{\rm{max-ex}}>\omega_{\rm{sl-ex}}$ implies
\begin{equation}
\frac{1}{\epsilon^{\prime}}\left[(1+\epsilon^{\prime})\sqrt{\alpha^2 +  \epsilon^{\prime^2} + 2\epsilon^{\prime}} - \alpha \right]< 2\alpha
\end{equation}
We proceed similarly to find the critical value of $\alpha$ in the extremal case.
\begin{equation}
\alpha^2> \frac{5\epsilon^{\prime}+ 2}{3 \epsilon^{\prime} +2} \equiv \alpha^2_{\rm{cr-ex}}
\label{condiex}
\end{equation}
If we perturb an extremal KTN space-time with a scalar field, and the initial parameters of the space-time satisfy (\ref{condiex}), the horizon can be destroyed, provided that the energy of the incoming field is $\delta E\sim M\epsilon^{\prime}$ and the frequency is in the range $\omega_{\rm{sl-ex}}<\omega<\omega_{\rm{cr-ex}}$. For $\alpha<\alpha_{\rm{cr-ex}}$, i.e. (\ref{condiex}) is not satisfied,  $\omega_{\rm{cr-ex}}$ is less than $\omega_{\rm{sl-ex}}$ so that an interval of frequencies that can destroy the horizon does not exist. As we discussed in the nearly extremal case, the extremal KTN black holes that do not satisfy (\ref{condiex}), cannot be destroyed by scalar fields.

For a numerical example let us consider $\epsilon^{\prime}=0.01$. In that case $\alpha_{\rm{cr-ex}}=1.0049$ up to four significant digits. If we take $\alpha=1<\alpha_{\rm{cr-ex}}$, we find that $\omega_{\rm{max-ex}}=0,4975 (m/M)$, and $\omega_{\rm{sl-ex}}=0,5 (m/M)$. One cannot find a frequency $\omega$ to destroy the horizon. On the other hand if we take $\alpha=1,01>\alpha_{\rm{cr-ex}}$, $\omega_{\rm{max-ex}}=0,4975 (m/M)$, and $\omega_{\rm{sl-ex}}=0,4950 (m/M)$ up to four significant digits, and the frequencies in the range $\omega_{\rm{sl-ex}}<\omega <\omega_{\rm{max-ex}}$ can be used to destroy the horizon. Let us consider an extremal KTN spacetime with $\alpha=1,01$, and perturb it with a scalar field. For this spacetime $J=M^2 \alpha =1.01 M^2$, and $\ell^2=M^2 \beta^2=M^2(\alpha^2 -1)=0.0201 M^2  $. By definition, the initial parameters of the spacetime satisfy $\delta_{\rm{in}}=M^2 + \ell^2 - J^2/M^2=0$. There exists a single event horizon at $r=M \pm \sqrt{\delta}=M$. After the interaction with the scalar field, the perturbed parameters satisfy
\begin{eqnarray}
\delta_{\rm{fin}} &=& (M+ \delta E)^2 + \ell^2 - \frac{(J+ \delta J)^2}{(M+ \delta E)^2 } \nonumber \\
&=& M^2\left[ (1+0.01)^2 + 0.0201 - \frac{(1.01+ 0.0202)^2}{(1+0.01)^2 } \right]\nonumber \\ &=& -0.0002 M^2
\label{ex4}
\end{eqnarray}
The negative result in (\ref{ex4}) indicates that the event horizon has been destroyed in the interaction of the extremal KTN spacetime with the scalar field. 
 
Previous works on the interactions of test fields with Kerr, Kerr-Newman, Kerr-Newman anti de Sitter, and BTZ spacetimes agree on the fact that scalar fields cannot destroy the event horizon if the spacetime is initially extremal~\cite{semiz,toth1,overspin,natario,btz}. From that point  of view, the destruction of the horizon in the KTN spacetime may  be unexpected. The question whether this contradicts the previous results may also arise, considering the fact that one obtains the Kerr spacetime in the limit   $\ell \to 0$. To avoid any confusion, we should note that scalar fields cannot destroy the horizon in  all extremal KTN spacetimes. The extremal KTN spacetime should also satisfy   the condition (\ref{condiex}), which implies that $\alpha^2_{\rm{cr-ex}}>1$ independent of the choice of $\epsilon^{\prime}$. In the limit $\ell \to 0$, this is equivalent to demanding $a^2>M^2$ for the initial values of the space-time parameters. Thus, the extremal KTN spacetimes that can be destroyed by scalar fields correspond to naked singularities in the Kerr limit. In this sense, there is no contradiction  with the previous results  for scalar fields interacting with the Kerr family of spacetimes.  

If we perturb the extremal space-time with neutrino fields instead, all the frequencies $\omega<\omega_{\rm{max-ex}}$ can be used to destroy the horizon, since superradiance does not occur for neutrino fields. That is a generic destruction as  in the case of the nearly extremal space-time. 
\section{Summary and conclusions}
In the previous studies with test fields, it was shown that scalar fields cannot destroy the horizon of an extremal black hole in Kerr and BTZ spacetimes. However, one can overspin a nearly extremal black hole into a naked singularity using tailored wave packets in a narrow range of frequencies. Since this range is bounded below by the superradiant limit, there is no lower bound for fermionic fields. Therefore fermionic fields can lead to a generic destruction of the horizon which also applies to extremal black holes~\cite{overspin,superrad,duztas,btz}. This result was generalized by Natario, Queimeda, and Vicente who proved that test fields satisfying the null energy condition cannot destroy the event horizon in Kerr-Newman and Kerr-Newman-anti de Sitter spacetimes~\cite{natario}. 

In this work we have investigated the possibility to destroy the event horizon in the interaction of the KTN space-time with test, massless scalar and neutrino fields. We constructed a thought experiment in which massless fields that are incident from infinity scatter off the KTN black hole. The main difference between scalar and neutrino fields is the occurrence of superradiance. 

For scalar fields we showed that the horizon can be destroyed if we start with a nearly extremal spacetime.  Analogous to the Kerr case one has to use tailored wave packets in a narrow range of frequencies bounded below by the superradiance limit. In the Kerr spacetime, such a range exists for all nearly extremal black holes. However, in the KTN spacetime, the initial angular momentum parameter should be sufficiently large relative to mass and NUT parameters for such a range to exist. The spacetime can be very close to extremality, but it will not be possible to destroy the horizon by sending in scalar fields from infinity unless the condition (\ref{condi2}) is satisfied. For that reason we start with a nearly extremal spacetime  that  satisfies (\ref{condi2}). Then we perturb this spacetime by sending in scalar fields from infinity with a frequency in the range $\omega_{\rm{sl}}<\omega<\omega_{\rm{cr}}$ and energy $\delta E \sim M\epsilon$. At the end of the interaction the final parameters of the spacetime do not satisfy (\ref{crit}), i.e. $\delta_{\rm{fin}}<0$.  The event horizon has been destroyed in the interaction of the spacetime with the scalar field. Unlike the Kerr case, it is not possible to overspin all nearly extremal KTN spacetimes. The procedure only applies to the nearly extremal KTN spacetimes with sufficiently large angular momentum.

We also derived that scalar fields can destroy extremal KTN black holes  with sufficiently large angular momentum relative to their mass and NUT charge. This is the first example in which a classical field  satisfying the null energy condition can actually destroy an extremal black hole. We derived the lower limit for the angular momentum parameter in (\ref{condiex}). We argued that, due to the condition (\ref{condiex}), the extremal KTN spacetimes that can be destroyed by scalar fields correspond to naked singularities in the Kerr limit $\ell \to 0$. This is in accord with the fact that extremal Kerr black holes cannot be destroyed by scalar fields. In \cite{natario}, it was also proved that test fields satisfying the null energy condition cannot destroy the event horizon in Kerr-Newman and Kerr-Newman-anti de Sitter spacetimes. The reason one cannot extend this result to the KTN spacetime is the fact that it is not causally well behaved.

It is known that superradiance does not occur for neutrino fields in the KTN spacetime. This leads to a generic destruction of the event horizon both for extremal and nearly extremal cases, analogous to the Kerr case.

In this analysis we neglected the backreaction effects. Naively, one can assume that backreaction effects can compensate for the destruction of the event horizon derived in this work, as  it does in the case of particles. In principle backreaction effects can be employed to fix the destruction of the horizon for the case of scalar fields. In this case the frequencies to destroy the horizon are restricted to a narrow range due to the occurrence of superradiance. For a nearly extremal spacetime with $\delta_{\rm{in}} \sim M\epsilon^2$, we showed that $\delta_{\rm{fin}}  \sim -M\epsilon^2$ after the interaction with the scalar field. In principle backreaction effects can be employed for this interaction to fix the negative results in (\ref{ex2}) and (\ref{ex4}), so that   $\delta_{\rm{fin}} \gtrsim 0$ and the horizon is stable.
However, in the case of neutrino fields all the modes in the range $0<\omega<\omega_{\rm{cr}}$ can be used to destroy the horizon. Backreaction effects become negligible as $\omega$ decreases without bound, and the relative contribution of the test field to the angular momentum of the spacetime compared to its contribution to the energy of the spacetime ($\delta J /\delta E$) increases without bound. For example, we considered a neutrino field with frequency $0.25(m/M)$. We derived that the absolute value of $\delta_{\rm{fin}}$ is approximately 400 times larger than the corresponding value in the thought experiment with the scalar field. We  can lower  the frequency even further to obtain a larger absolute value for $\delta_{\rm{fin}}$. Backreaction effects can only compensate for the destruction of the horizon with $\delta_{\rm{fin}} \sim -M\epsilon^2$. They become negligible as we lower the frequency slightly below the superradiant limit, far before we reach $(\omega_{\rm{sl}}/2) \sim 0.25(m/M)$.  In this respect the destruction of the event horizon by neutrino fields is generic, and it cannot be fixed by backreaction effects. This is analogous to the previous results for neutrino perturbations of Kerr and BTZ spacetimes~\cite{superrad,duztas,toth,mode,btz}.

In addition to the backreaction effects, one can also argue that the astrophysical limitation on the value of $a/M$ can invalidate the calculations in this work.  Thorne derived that the value of  $a/M$ cannot exceed 0.998 for Kerr black holes swallowing matter and radiation from an accretion disk~\cite{thorne}. A similar limit can also exist for  KTN black holes interacting with massless fields. This will bring a lower limit to  $\epsilon$ which parametrizes the black hole’s closeness to extremality. We can adjust the energy of the incoming field so that the black hole loses its event horizon. However, if we increase the energy of the incoming field above a certain value, the test field  approximation can be violated, and the calculations will not be valid. The problem is that  there is no definite criteria agreed upon,  for the magnitude of the energy and angular momentum of a test field relative to the background spacetime. Under these circumstances, we can say that the calculations are valid to the extent that the test field approximation can be considered reasonable.

\end{document}